\documentclass[10pt]{iopart}
\usepackage{graphicx,rotating,subfigure} 
\usepackage{iopams}  
\usepackage{dsfont}
\usepackage{psfrag}

\usepackage{graphicx}
\usepackage{dcolumn}
\usepackage{bm}
\usepackage{epsfig}
\newcommand{\be}{\begin{eqnarray}}
\newcommand{\ee}{\end{eqnarray}}
\newcommand{\xop}{\mathbf{b}}
\newcommand{\yop}{{\mathbf{b}^\dagger}}
\newcommand{\Nop}{\mathbf{n}}

\def\B{\mathfrak B}
\def\BB{{\overline{{\mathfrak{B}}}}}
\def\ba{\mathfrak b}
\def\bb{{\overline{{\mathfrak{b}}}}}
\def\refeq#1{(\ref{#1})}
\def\d{\mbox d}

\def\nn{\nonumber}
\def\i{\int_{-\infty}^{\infty}}

\def\b{\beta}

\def\Or{\mathcal O}
\def\g{\gamma}
\def\al{\alpha}

\def\e{\epsilon}

\def\ve{\varepsilon}
\def\l{\left}
\def\r{\right}
\def\te{\mbox{e}}
\def\rmi{{\rm i}}

\begin{document}
\bibliographystyle{jpa}
\title{Low-temperature asymptotics of integrable systems in an external field}
\author{Michael Bortz}
\address{Department of Theoretical Physics, Research School of Physics and Engineering, Australian National University, Canberra ACT 0200, Australia}
\begin{abstract}
An asymptotic low-temperature expansion is performed for an integrable bosonic
lattice model and for the critical spin-1/2 Heisenberg chain in a magnetic field. The results
apply to the integrable Bose gas as well. We also comment on a high-temperature expansion of the bosonic lattice model.  
\end{abstract}
\section{Introduction}
The calculation of the relevant thermodynamic potential of an interacting
many-body system in the thermodynamic limit belongs to the most fundamental
problems of statistical mechanics. For Yang-Baxter integrable models, very
sophisticated techniques like the thermodynamic Bethe ansatz (TBA; for a
review, see \cite{tak99}) and the quantum transfer matrix (QTM; for a review,
see \cite{ak04}) have been developed which yield exact expressions for the
thermodynamic potential in terms of certain auxiliary functions. These
auxiliary functions are determined by non-linear integral equations, which are
well suited for numerical investigations (see, for example,
\cite{yan70,ak98,ak00}). 

As far as the spin-1/2 $XXZ$-Heisenberg chain is concerned, high-temperature expansions have been
performed analytically (see, for example, \cite{tak99,shi02,tsu03}) up to
high orders. At low temperatures, the leading (``universal'') terms have been
obtained from the exact solution. In the absence of a magnetic field, higher orders were calculated by combining the $T=0$ Bethe ansatz with an
effective field theory for the low-lying excitations \cite{luk98,luk03}. It
still remains a challenging problem to perform such a low-temperature
expansion directly from the exact solution for the thermodynamical potential
at finite temperatures. 

On the other hand, for integrable bosonic systems, only the ``universal'' terms
stemming from the Luttinger-liquid character are known for general coupling
and particle density \cite{caz04,borser06}. A virial expansion of the
integrable Bose gas was given in \cite{yan70}. Very recently, the low-temperature
behaviour of an anyonic gas at large interaction strength was considered
\cite{bat06}. However, a systematic low-temperature expansion for integrable
bosonic systems with arbitrary interaction and occupation has not been performed yet.

In this article, a Sommerfeld-type expansion is applied to the
TBA-/QTM-equations for integrable systems in an external field, namely the
$q$-Bose model with a chemical potential and the critical $XXZ$-chain in a
magnetic field. The results are directly transferable to the integrable Bose
gas, and the method itself can be employed for other integrable systems with
external fields. The importance of this work lies in the close connection
between the low-energy effective Hamiltonian and the low-temperature expansion
of the thermodynamical potential, as pointed out in \cite{luk98,luk03} for the
zero-field case of the Heisenberg chain. The low-lying excitations of this
model, as well as of the $q$-Bose model, are described by a Gaussian model. In
\cite{luk98,luk03}, the leading irrelevant operators and its coefficients were
identified for the Heisenberg chain in zero field. For finite field, the leading irrelevant operators have been found
recently \cite{aff06}. It is an interesting problem now to relate the
low-temperature expansion found here to the coefficients of the irrelevant
operators in this effective Hamiltonian for the spin chain. Besides that, for
the $q$-Bose model, only
the scaling dimension of the leading irrelevant operators are known, but not
the operators themselves nor the coefficients. This work constitutes a first
step to fill this gap. The knowledge of the effective Hamiltonian allows for
the calculation of asymptotics of correlation functions, \cite{luk98,luk03,aff06}.    

Our main result is an asymptotic low-temperature expansion of the thermodynamic
potential $g=g(T,\mu)$ in even powers of the temperature, 
\be
g(T,\mu)=\sum_\nu
g_\nu(\mu)T^{2\nu},\qquad \nu=0,1,2,\ldots \;.\label{geng}
\ee
The coefficients $g_{\nu}(\mu)$
depend on the chemical potential $\mu$ only. For the Heisenberg chain, $\mu$
is replaced by the magnetic field $h$. We emphasize that Eq.~\refeq{geng} is
not a Taylor expansion in $T$, but rather an asymptotic expansion. This means
that the coefficient in order $T^{2 m}$ is defined as
\be
\lim_{T\to0} \frac{g(T,\mu)-\sum_{\nu=0}^{m-1} g_\nu(\mu) T^{2 \nu}}{T^{2
    m}}=:g_m(\mu)\nn.
\ee
The first two coefficients in Eq.~\refeq{geng} are well known:
$g_0$ constitutes the ground state energy. Furthermore, $g_1=-\pi/(6 v)$, where
$v$ is the velocity of elementary excitations (``universal term''). Our approach allows for the
calculation of $g_{\nu\geq 2}$, and the calculation of $g_2$ is performed
explicitly in this article. 

The universal term originates in the Gaussian
part of the effective Hamiltonian, with scaling dimension two. The
higher-order terms stem from irrelevant operators with scaling dimensions
$2\nu$ ($\nu\geq 2$). This is consistent with the absence of backscattering
processes in both the $q$-Bose model and the Heisenberg chain in a magnetic
field: For the $q$-Bosons, the conserved particle current excludes
backscattering \cite{bor06}. On the other hand, a finite magnetic field
applied to the Heisenberg chain moves the system away from half filling (in
the fermionic picture) and thus forbids backscattering at low energies as
well. 

This article is organized as follows. In the next section, the low-temperature
expansion of $g$ for the $q$-Bose model is
obtained from the corresponding TBA-equations. The results are checked for a
certain choice of the interaction parameter (the phase-model,
\cite{bog97,bog98}) by comparing with an independent expansion in this case
and with numerical results. The third section deals with a high-temperature
expansion of the $q$-Boson model, whereas the fourth section contains the
low-temperature expansion for the critical spin-1/2 $XXZ$-chain. Calculations
not directly necessary for the understanding of the main text are deferred to
the appendix. In all calculations, units are chosen such that Boltzmann's constant $k_B\equiv 1$.     

\section{Low-temperature asymptotics of the $q$-Bose model}
\subsection{Definition of the model}
The $q$-Bose model is a one-dimensional integrable bosonic lattice model that
in an appropriate continuum limit leads to the Lieb-Liniger Bose gas. The
lattice regularization stems from a $q$-deformation of the underlying
commutators \cite{bog92,bog93,bul95}. Very recently, ground state and
thermodynamic properties have been calculated numerically and the Gaussian
part of the effective Hamiltonian has been identified
\cite{borser06,bor06}. Let us shortly review the definition of the model. 

$q$-deformed bosonic annihilation and creation operators are defined as
\be
\xop\yop=1-q^{2\Nop+2}\;,\quad \yop\xop = 1-q^{2\Nop}\;,\nn\\
\quad \xop
\Nop=(\Nop+1)\xop\;,\quad \yop \Nop=(\Nop-1)\yop\;,\nn
\ee
with $0\leq q<1$. In the following, it is often convenient to use the parameterization
\be
q=\te^{-\eta},\; \eta\in\l]0,\infty\r[\label{qeta}.
\ee
A $q$-oscillator is represented in the Fock space:
\begin{equation}\label{Fock}
\Nop|n\rangle=n|n\rangle\;; \quad n = 0,1,2,3,\dots\;;\;
\xop|0\rangle =0\;.
\end{equation}
We deal with a chain of length $L$ where the local $q$-oscillator
algebra is assigned to each site $\ell=1,\ldots,L$. The quantities 
\be
\mathcal{N}&=&\sum_{\ell=1}^L \Nop_\ell\nn\\
\mathcal{P}_+&=&\frac{1}{1-q^2}\sum_{\ell=1}^L \xop_{\;\ell\;}^{} \yop_{\;\ell+1\;}^{}\;,\mathcal{P}_-=\frac{1}{1-q^2} \sum_{\ell=1}^L
\xop_{\;\ell\;}^{} \yop_{\;\ell-1\;}^{}\;,\nn
\ee
commute pair-wisely. The eigenvalue of $\mathcal N$ is the number of particles $N$. The choice of the Hamiltonian
\be
\mathcal{H}\;=\;- \frac{1}{2} (\mathcal{P}_+ + \mathcal{P}_-)-\mu
\mathcal{N}\,, \label{defh}
\ee
where $\mu$ is the chemical potential, leads \cite{bog92,bog93,bul95} in the continuum limit to the Lieb-Liniger Bose gas with Hamiltonian
\be
\fl \mathcal H= \frac{\Delta}{2} \int \l[ \partial_x \Psi^\dagger(x) \partial_x \Psi(x)+  2c \Psi^\dagger(x)\Psi^\dagger(x)\Psi(x)\Psi(x)-\mu\Psi^\dagger(x)\Psi(x)\r]\d x \label{nls}.
\ee
Here $\Delta$ is the lattice constant of the $q$-Bose model and the coupling constant $c$ has been introduced such that $\eta=c\cdot \Delta$. Furthermore, trivial constants have been absorbed into $\mu$. 

Using the TBA, the thermodynamic potential potential (which here is $-p$, with
$p$ being the pressure) was found to be given by \cite{bul95}
\be
g(T,\mu)&=& -T \int_{-\pi}^\pi \ln \l( 1+ \te^{-\ve(k)/T}\r) \,\frac{\d k}{2\pi}\label{g},
\ee
where the function $\ve(k)$ is obtained from
\be
\fl\ve(k)= -\cos k -\mu -T \int_{-\pi}^\pi \frac{\sinh 2\eta}{\cosh 2\eta - \cos(k-k')}\, \ln\l(1+\te^{-\ve(k')/T}\r) \,\frac{\d k'}{2\pi}\,.\label{vefint}
\ee
These two equations are valid for arbitrary $T$. They constitute the starting
point for our low-temperature analysis. We will calculate contributions to $g$ including $\Or(T^4)$. However, higher orders can be calculated as well in the approach sketched below.   

\subsection{Low-temperature expansion}
At $T=0$, the function $\ve(k)|_{T=0}$ has two zeroes $\pm B(T=0)=:\pm B_0$. In \cite{borser06}, the ranges of the particle density $n:=N/L$, the chemical potential $\mu$ and the parameter $B_0$ were given as in table \ref{tab1}. 
\begin{table}[h]
\begin{center}
\begin{tabular}{c|cc}
$n$ & 0 & $\infty$ \\
\hline
$\mu$ & -1 & 0 \\
\hline
$B_0$ & 0 & $\pi$
\end{tabular}
\caption{The ranges of the parameters $n,\mu,B_0$.} 
\label{tab1}  
\end{center} 
\end{table} 
Here, we are dealing with temperatures $T\ll |\mu|$, i.e. $\beta:=1/T\gg |\mu|$, and we take the $T=0$-case as a point of reference for the ranges $n,\mu,B$ in the low-temperature regime. To avoid possible singularities, we assume a finite non-zero density $0<n<\infty$, i.e. $-1<\mu<0$, $0<B<\pi$. Thus the two zeroes of $\ve$ are supposed to lie inside the interval $\l]-\pi,\pi\r[$.

Since $\ve(|k|>B)>0$, terms $\sim \exp\l[-\beta \ve(\pm \pi)\r]$ are exponentially small. This allows us to eliminate exponentially small quantities by performing two integrations by parts in Eq.~\refeq{g}:
\numparts
\label{ip}
\be
\fl g(T,\mu)= \int_{-\pi}^\pi \frac{\ve'(k)k}{1+\te^{-\b\ve(k)}}\frac{\d k}{2\pi} -T \l.\ln\l(1+\te^{\beta \ve(k)}\r)\frac{k}{2\pi}\r|_{-\pi}^\pi\label{ip1}\\
\fl\qquad=- \beta\int_{-\pi}^\pi \frac{\l.\ve(k')k'\r|_k-\int_k\ve(k')\,\d k'}{4 \cosh^2\frac{\b\ve(k)}{2}}\,\ve'(k)\frac{\d k}{2\pi}+\l.\frac{1}{2\pi}\,\frac{\ve(k) k -\int_k\ve(k')\d k'}{1+\te^{-\beta \ve(k)}}\r|_{-\pi}^\pi\label{ip2}\\
\fl\qquad=- \i \frac{\beta^{-1}\l. u\,k(u)\r|_{k_-(u)}^{k_+(u)}- \int_{k_-(u)}^{k_+(u)}\ve(k')\d k'}{4\cosh^2\frac{u}{2}}\, \frac{\d u}{2\pi}\label{intp}.
\ee
\endnumparts
As explained above, the boundary terms stemming from the integrations by parts
are exponentially small and are therefore neglected. In the last line,
$u=\beta \ve(k)$ was substituted and the integration boundaries sent to $\pm \infty$, which results again only into exponentially small corrections. Since $u/\beta$ is small by construction, we Taylor-expand $\ve(k)$ around $\pm B$ to determine the boundaries $k_\pm$ for the integrand in Eq.~\refeq{intp}:
\be
\fl\frac{u}{\b}=(k_+-B) \ve'(B) + \frac12(k_+-B)^2\ve''(B) + \frac16(k_+-B)^3 \ve^{(3)}(B)+\Or(k_+^4)\nn\\
\fl k_+=B+\frac{u}{\beta \ve'(B)}-\frac{\ve''(B)}{2\ve'^3(B)}\frac{u^2}{\beta^2}+\frac{3(\ve''(B))^2-\ve'(B)\ve^{(3)}(B)}{6 \ve'^5(B)}\frac{u^3}{\beta^3}+\Or\l(\beta^{-4}\r)\nn.
\ee
Analogously, 
\be
\fl k_-=-B-\frac{u}{\beta \ve'(B)}+\frac{\ve''(B)}{2\ve'^3(B)}\frac{u^2}{\beta^2}-\frac{3(\ve''(B))^2-\ve'\ve^{(3)}(B)}{6 \ve'^5(B)}\frac{u^3}{\beta^3}+\Or\l(\beta^{-4}\r)\nn.
\ee
We now expand the integrand of the outer integral in Eq.~\refeq{intp} in powers of $u$. Only even powers contribute, leading to integrals of the type
\numparts
\label{zeta}
\be
\i \frac{1}{4\cosh^2\frac{u}{2}}\,\d u &=&1\label{zeta1}\\
\i \frac{u^{2 n}}{4\cosh^2\frac{u}{2}}\,\d u &=& (2n-1)!4n \l(1-2^{1-2 n}\r)\zeta(2 n)\,,\; n=1,2,\ldots\label{zeta2}
\ee
\endnumparts
Especially, $\zeta(2)=\pi^2/6$, $\zeta(4)=\pi^4/90$. Then
\be
\fl g(T,\mu)=  \int_{-B}^B\ve(k')\frac{\d k'}{2\pi} - \frac{\pi}{6\ve'(B)} T^2 - \frac{7 \pi^3}{30} \frac{3(\ve''(B))^2-\ve^{(3)}(B)\ve'(B)}{12 \ve'^5(B)}T^4+\Or\l(T^{6}\r).\label{bg}
\ee
In order to identify the $T$-dependence of $\ve(k)$ and $B$, one performs an asymptotic low-temperature expansion of the convolution in Eq.~\refeq{vefint} in an analogous way:
\be
\fl\int_{-\pi}^\pi\kappa(k-k') \ln\l(1+\te^{-\ve(k')/T}\r) \,\frac{\d k'}{2\pi}= -\b\int_{-B}^B  \kappa(k-k')\ve(k')\frac{\d k'}{2\pi}\nn\\
\fl\qquad +T\frac{\pi}{12} \frac{\kappa(k-B)+\kappa(k+B)}{ \ve'(B)}\nn\\
\fl\qquad +T^3\frac{7\pi^3}{60}\l\{\l(\frac{(\ve''(B))^2}{4\ve'^5(B)}-\frac{\ve^{(3)}(B)}{12\ve'^4(B)}\r)\l[\kappa(k-B)+\kappa(k+B)\r]-\frac{\ve''(B)}{4\ve'^4(B)}\r.\nn\\
\fl\qquad\;\;\times\l[\kappa'(k-B)+\kappa'(k+B)\r]+\l.\frac{\l[\kappa''(k-B)+\kappa''(k+B)\r]}{12 \ve'^3(B)}\r\}\nn.
\ee
Thus the integral equation for $\ve(k)$ takes the following form:
\be
\ve(k)=\ve^{(0)}(k) + \int_{-B}^B  \kappa(k-k')\ve(k')\frac{\d k'}{2\pi}\label{eps}
\ee
where (we drop the neglected $\Or\l(T^6\r)$ in the following) 
\be
\fl\ve^{(0)}(k)= -\cos k -\mu+\int_{-B}^B\kappa(k-k')\ve(k')\frac{\d k'}{2\pi}-T^2\frac{\pi}{12} \frac{\kappa(k-B)+\kappa(k+B)}{ \ve'(B)}\nn\\
\fl\qquad -T^4\frac{7\pi^3}{60}\l\{\l(\frac{(\ve''(B))^2}{4\ve'^5(B)}-\frac{\ve^{(3)}(B)}{12\ve'^4(B)}\r)\l[\kappa(k-B)+\kappa(k+B)\r]-\frac{\ve''(B)}{4\ve'^4(B)}\r.\nn\\
\fl\qquad\;\;\times\l[\kappa'(k-B)+\kappa'(k+B)\r]+\l.\frac{\l[\kappa''(k-B)+\kappa''(k+B)\r]}{12 \ve'^3(B)}\r\}\label{ve0}.
\ee
Consider now the function $\rho(k)$, defined by
\be
\rho(k)&=&\rho^{(0)}(k) + \int_{-B}^B\kappa(k-k') \rho(k') \frac{\d k'}{2\pi}\label{rho}\\
\rho^{(0)}(k)&=&\frac{1}{2\pi}\label{rho0}.
\ee
Note that this function is equal to the density of Bethe ansatz roots in the thermodynamic limit only for $T=0$. For $T>0$, the integration boundary $B$ acquires a $T$-dependence, and so does $\rho$. 

Due to the symmetry $\kappa(k)=\kappa(-k)$, the equality $\int_{-B}^B \rho^{(0)}(k)\ve(k)\d k = \int_{-B}^B\rho(k)\ve^{(0)}(k)\d k$ holds. Using this relation and Eq.~\refeq{ve0}, we can manipulate Eq.~\refeq{bg} further and obtain
\be
\fl g(T,\mu)=\int_{-B}^B\rho(k)\l(-\cos(k)-\mu\r)\d k - T^2 \frac{\pi^2}{3} \frac{ \rho(B)}{\ve'(B)}\nn\\
\fl\qquad - T^4\frac{7\pi^4}{15} \l[ \rho(B)\l(\frac{1}{4} \frac{(\ve''(B))^2}{\ve'^5(B)} - \frac{\ve^{(3)}(B)}{12\ve'^4(B)}\r)-\frac{\rho'(B)\ve''(B)}{4\ve'^4(B)} + \frac{\rho''(B)}{12\ve'^3(B)}\r]\label{gt}.
\ee
We now have to identify the $T$-dependence of $\rho(k)$, $\ve(k)$ and $B$. Since in the equations for $\rho(k)$ and $\ve(k)$, only even powers of $T$ occur, we make the $T$-dependence explicit as follows:
\numparts
\label{tdep}
\be
B&=:&B_0+B_1 T^2 + B_2 T^4\label{b}\\
\rho(k)&=:& \rho_0(k)+\rho_1(k)T^2 + \rho_2(k)T^4\label{rhot}\\
\ve(k)&=:& \ve_0(k) +\ve_1(k)T^2 +\ve_2(k)T^4\label{epst}.
\ee
\endnumparts
Let us consider Eq.~\refeq{eps} first. After inserting (\ref{b}, \ref{epst}), one obtains equations that determine $\ve_\nu$, $\nu=0,1$:
\numparts
\label{veint}
\be
\fl\ve_0(k)=-\cos k -\mu +\int_{-B_0}^{B_0}\kappa(k-k')\ve_0(k')\frac{\d k'}{2\pi}\label{vet0}\\
\fl\ve_1(k)= -\frac{\pi}{12\ve_0'}\l[\kappa(k-B_0)+\kappa(k+B_0)\r] +\int_{-B_0}^{B_0}\kappa(k-k')\ve_1(k')\frac{\d k'}{2\pi}\label{vet1}
\ee
\endnumparts
where we abbreviated $\ve'_0(B_0)=:\ve'_0$. Obviously, $B_1$ is given through
\be
B_1=-\frac{\ve_1(B_0)}{\ve'_0(B_0)}\label{b1}.
\ee
Note that $B_1>0$. It will become clear below that we do not need $\ve_2(k)$. Analogously, equations for $\rho_{1,2}$ are derived:
\numparts
\label{rhonu}
\be
\rho_\nu(k)&=&\rho^{(\nu)}(k) + \int_{-B_0}^{B_0} \kappa(k-k') \rho_\nu(k')\frac{\d k'}{2\pi}\label{rhonuex}\\
\rho^{(1)}(k)&=&B_1\frac{\rho_0}{2\pi}\l[\kappa(k-B_0)+\kappa(k+B_0)\r]\label{rho1}\\
\rho^{(2)}(k)&=&\frac{1}{2\pi}\l(B_2\rho_0+B_1\rho_1+\frac12B_1^2\rho_0'\r)\l[\kappa(k-B_0)+\kappa(k+B_0)\r]\nn\\
& & +\frac{1}{4\pi}B_1^2\rho_0\l[\kappa'(k-B_0)-\kappa'(k+B_0)\r]
\ee
\endnumparts
where $\rho_\nu(B_0)=:\rho_\nu$ etc. (see Eq.~\refeq{rho0} for $\rho_0$). 

On the other hand, with $d_\ve(k):=-\cos k -\mu$, 
\be
\fl\int_{-B}^B\rho(k)\l[-\cos k-\mu\r]\d k=\int_{-B_0}^{B_0}\rho_0(k)d_\ve(k)\d k\nn\\
\fl\;\; + T^2\l\{\int_{-B_0}^{B_0}\rho_1(k)d_\ve(k)\d k +2 B_1\rho_0d_\ve(B_0)\r\}\label{int1}\\
\fl\;\;+T^4\l\{\int_{-B_0}^{B_0}\rho_2(k)d_\ve(k)\d k+2\l[ B_2\rho_0+B_1\rho_1+\frac12 B_1^2\rho_0'\r]d_\ve(B_0)+B_1^2 \rho_0\,d_\ve'(B_0)\r\}\nn.
\ee
We insert now \refeq{rhonu} into Eq.~\refeq{int1} and make use of $\ve_0(B_0)=0$.
One ends up with
\be
\fl\int_{-B}^B\rho(k)\l[-\cos k-\mu\r]\d k=\int_{-B_0}^{B_0}\rho_0(k)\l[-\cos k-\mu\r]\d k+T^4 \rho_0 \ve_0' B_1^2\label{con1}.
\ee
Let us look at the second term on the rhs of Eq.~\refeq{gt}. Here, we
identify a $T^4$-contribution as well, originating in:
\be
\frac{\rho(B)}{\ve_0'(B)}=\frac{\rho_0}{\ve_0'} + T^2\l[\frac{\rho_1}{\ve'_0}-\frac{\ve_1'\rho_0}{\ve_0'^2}+B_1\l(\frac{\rho_0'}{\ve_0'}-\frac{\ve_0''\rho_0}{\ve_0'^2}\r)\r]\label{con2},
\ee
where, again, $\ve_\nu(B_0)=:\ve_\nu$ etc. In the last equation, it is convenient to represent $\rho_1$ in terms of $\ve_1$, which follows from Eqs.~(\ref{vet1}, \ref{rhonuex}, \ref{rho1}): 
\be
\rho_1(k)=-\frac{6}{\pi^2}B_1 \rho_0 \ve_0'\,\ve_1(k)\label{rhove}. 
\ee
Taking Eqs.~(\ref{gt}, \ref{con1}, \ref{con2}, \ref{rhove}) together, we arrive at the low-temperature expansion of $g$:
\numparts
\label{gseries}
\be
g(T,\mu)&=&g_0(\mu)+g_1(\mu)T^2+g_2(\mu)T^4+\Or\l(T^6\r)\label{ggen}\\
g_0(\mu)&=&\int_{-B_0}^{B_0}\rho_0(k)\l(-\cos k-\mu\r)\frac{\d k}{2\pi}\label{g0ex}\\
g_1(\mu)&= & -\frac{\pi^2}{3} \frac{2\pi\rho_0}{\ve_0'}\label{g1ex}\\
g_2(\mu)&= & -\l\{\frac{7\pi^4}{15}\l[\rho_0\l(\frac{(\ve_0'')^2}{4\ve_0'^5}-\frac{\ve_0^{(3)}}{12\ve_0'^4}\r)-\frac{\rho_0'\ve_0''}{4\ve_0'^4}+\frac{\rho_0''}{12\ve_0'^3}\r]\r.\nn\\
& &\l.+\frac{\rho_0\ve_1^2}{\ve_0'}+\frac{\pi^2}{3}\l[-\frac{\ve_1'\rho_0}{\ve_0'^2}-\frac{\rho_0'\ve_1}{\ve_0'^2}+\frac{\ve_0''\rho_0\ve_1}{\ve_0'^3}\r]\r\}\label{g2}
\ee
\endnumparts
The order $T^2$ has been obtained previously \cite{borser06} and is common to one-dimensional critical systems \cite{korbook}: The velocity $v_c$ associated with the low-lying excitations reads $v_c=\ve_0'/(2\pi \rho_0)$, so that $g_1(\mu)=\frac{\pi}{6 v_c}$. Let us shortly comment on $g_2$. In the first line of Eq.~\refeq{g2}, only $\ve_0',\rho_0$ and derivatives of these functions taken at $B_0$ enter. All these quantities are obtained from the $T=0$ Bethe ansatz, analogously to $v_c$. Additionally, the second line of Eq.~\refeq{g2} contains $\ve_1$, given through Eq.~\refeq{vet1}. 

From \refeq{gseries}, the charge susceptibility $\chi_c(T,\mu)=-\partial^2_\mu
g(T,\mu)$ as a function of $T$ and $\mu$ is derived. When calculating the
specific heat however, attention has to be paid that it is defined at constant particle density, whereas above, calculations have been carried out at constant chemical potential. Thus the thermodynamic potential acquires a $T$-dependence \cite{aktj97}
\be
\partial_T\mu|_n&=&-\frac{\partial_T n|_\mu}{\partial_\mu n|_T}\label{mut}
\ee
and we have to consider
\be
\l.\frac{C(T,n)}{T}\r|_n&=& -\partial^2_T g|_{\mu} - \frac{\l(\partial_T n|_\mu\r)^2}{\partial_\mu n|_T}\label{ct}
\ee
Analogously to (\ref{b}, \ref{rhot}, \ref{epst}), we write the $T$-dependence of $\mu(T,n)$ explicitly, 
\be
\mu(T,n)=\mu_0(n)+\mu_1(n)T^2\nn\,.
\ee
Then, according to Eq.~\refeq{mut}, 
\be
\mu_1(n)=-\frac{\pi}{6\chi_c(T=0,n)}\partial_{\mu_0}\frac{1}{v_c}=-\frac{\pi}{6}\partial_n\frac{1}{v_c}\label{mu1}\,,
\ee
where $\chi_c(T=0,n)=K_c/v_c$, with $K_c$ being the Luttinger parameter, to be obtained from $K_c=4 \pi \rho_0(B_0)$ \cite{borser06}. Note that the same symbols $\chi_c$, $v_c$ are being used for the susceptibility and velocity as functions of $\mu_0$ and of $n$. The $T^2$-dependence of $\mu$ leads to an additional contribution to the $T^2$-term in $C/T$, stemming from the first term on the rhs of Eq.~\refeq{ct}. The final result is 
\be
\fl\frac{C(T,n)}{T}= \frac{\pi}{3 v_c(\mu_0(n))}+T^2\l[-12 g_2(\mu_0(n))+\frac{\pi^2}{6}\l(\frac{\partial_{\mu_0} v_c}{v_c^2}\r)^2\frac{1}{\chi_c(T=0,n)}\r]\nn\\
\fl\qquad+\Or(T^4)\label{ctn}.
\ee

Let us shortly comment on the integrable Lieb-Liniger Bose gas,
Eq.~\refeq{nls}. The analogues of Eqs.~(\ref{g}, \ref{vefint}, \ref{rho}) read
\be
\rho(k)&=&\frac{1}{2\pi} +\int_{-B}^B \frac{2 c}{c^2+(k-k')^2} \rho(k') \frac{\d
  k'}{2\pi}\label{rhogas}\\
\ve(k)&=&k^2-\mu-T\i\frac{2 c}{c^2+(k-k')^2} \ln\l(1+\te^{-\beta \ve(k')}\r) \frac{\d
  k'}{2\pi}\nn\\
g(T,\mu)&=& -T\i \ln\l(1+\te^{-\beta \ve(k')}\r) \frac{\d
  k'}{2\pi}\nn\; .
\ee
Setting up equations for $\rho_{0,1}$, $\ve_{0,1}$ as above, the results
(\ref{g0ex}-\ref{g2}) yield $g_{0,1,2}$. The case of very strong interaction $c/n\gg 1$ allows for the explicit calculation of $g_{0,1,2}$ including the order $n/c$, because in this case, the integration kernel in Eq.~\refeq{rhogas} (and in the equations for $\ve_{0,1}$) becomes independent of the spectral parameter. As a result, 
\be
\rho_0(k)&=&\frac{1}{2\pi}\l(1+\frac{2B_0}{c\pi}\r),\qquad B_0=n \pi + \frac{2 n^2\pi}{c} \nn\\
\ve_0'(k)&=& 2 k,\qquad \ve_1(k)=-\frac{\pi}{6cB_0}\nn\\
g_0&=&-\frac{2n^3\pi^3}{3\pi}\l(1-\frac{6 n \pi}{c}\r)\nn\\
g_1&=&-\frac{1}{12n}-\frac{1}{3 c}\nn\\
g_2&=&-\frac{7}{960 \pi^2 n^5}-\frac{11}{1440\pi^2 n^4 c}\nn,
\ee
where $n\equiv n(\mu)$ with $\mu(n)=n^2\pi^2-\frac{16 n^3\pi^2}{3c}$. 
The chemical potential and the specific heat at constant density read
\be
\mu(T,n)&=&\l(n^2\pi^2-\frac{16 n^3\pi^2}{3c}\r)+\frac{T^2}{12 n^2}+\Or(T^4)\nn\\
\frac{C(T,n)}{T}&=&\l(\frac{1}{6 n}+\frac{2}{3 c}\r)+\l(\frac{1}{12 n^5\pi^2} -\frac{3}{40\pi^2n^4 c}\r)T^2\nn\; .
\ee 

\subsection{The $q=0$-case}
In this section, we focus on the case $q=0$ in Eq.~\refeq{qeta},
i.e. $\eta\to\infty$.  The corresponding model has been termed the ``phase-model'' in \cite{bog97,bog98}. This case is intriguing in so far as ground-state properties at $T=0$ can be calculated explicitly (i.e., the corresponding linear integral equations can be solved), while at the same time, the model still describes correlated bosonic particles. The reason for the solvability of the $T=0$-equations is that the integration kernel $\kappa(k)\equiv 1$ here. Since in the previous section, the coefficients of the low-temperature expansion have been expressed in terms of $T=0$-quantities only, these coefficients can be calculated explicitly for $q=0$. Furthermore, a very accurate numerical iteration procedure is possible in this case, which allows us to test the analytic results against the numerics. 

We defer the detailed calculations to the appendix. The specific heat at constant density reads:
\be
\fl\frac{C(T,n)}{T}=\frac{\pi}{3}\frac{n+1}{\sin\frac{n \pi}{n+1}} + \frac{\pi^3}{60} \frac{(n+1)\l(23+9\cos \frac{2 n \pi}{n+1}\r)}{\sin^5\frac{n\pi}{n+1}}T^2 \nn\\
\fl\qquad +\frac{\pi^5}{12096}\frac{(n+1)\l(33183+28012 \cos\frac{2 n \pi}{n+1}+1525 \cos\frac{4 n \pi}{n+1}\r)}{\sin^9 \frac{n\pi}{n+1}}T^4+\Or(T^6).\label{sphq0}
\ee
It is worthwhile noting that all coefficients are positive. Furthermore, for $n\to 0$, the coefficient of the order $T^{2\nu}$ diverges $\sim n^{-4\nu+1}$, whereas for $n\to \infty$, the divergence is $\sim n^{4\nu+2}$. Note that both divergencies are excluded from the beginning by having restricted $\mu$ to $-1<\mu<0$, that is, the density is supposed to be finite non-zero.

%
\section{Comment on the high-temperature expansion of the $q$-Bose model}
In this section, we indicate how to perform a high-temperature expansion of the $q$-Bose model at fixed density $n$. The general formulation is presented first, before concentrating on the $q=0$-case. The goal is an asymptotic expansion of the form $g(T,\mu)/T=\sum_\nu g_\nu(\mu)\beta^\nu$ (in a slight abuse of notation, we use the same symbols for the coefficients as in \refeq{geng}), with $\nu=0,1,2,\ldots$. 
\subsection{General formalism}
Let us write the integral equation Eq.~\refeq{vefint} in terms of $\ln a(k):=-\beta \ve(k)$ and $A=1+a$:
\be
\ln a(k)&=& \beta \cos k +\beta \mu + \int_{-\pi}^\pi \kappa(k-k') \ln A(k')\frac{\d k'}{2\pi}\label{aint}.
\ee
The $k$-dependent driving term on the rhs is bounded. Similarly to the low-temperature expansion, we make the $T$-dependence explicit in the following ansatz:
\numparts
\label{aht}
\be
\fl\ln A=\alpha_0 + \frac{\alpha_1}{T}+\frac{\alpha_2}{T^2}+\Or\l(T^{-3}\r)\\
\fl\ln a=\ln\l(\te^{\al_0}-1\r)+\frac{\al_1}{1-\te^{-\al_0}}\,T^{-1}+\frac{-\al_1^2+2\al_2\l(-1+\te^{\al_0}\r)}{4\cosh^2\frac{\al_0}{2}}\, T^{-2}.
\ee
\endnumparts
Since $n$ is to be held constant, we have to retain $\mu$ in Eq.~\refeq{aint} so that its $T$-dependence can be determined. Thus in zeroth order,
\be
\al_0=-\ln\l(1-\te^{\b \mu}\r)\nn\,.
\ee
Furthermore, from Eq.~\refeq{g} it follows for the particle density $n$ that
$n=\partial_{\beta \mu} \al_0$, which results in
\be 
\b \mu&=& \ln\frac{n}{n+1}\label{bmht}.
\ee
Here, the leading $T$-dependence of $\mu$ such that $n$ is constant is
displayed : $\mu\sim T$. From this, the canonical potential
$g_c:=g(T,\mu(T,n))+\mu(T,n) n$ is calculated:
\be
g_c&=&-T\l[ (n+1)\ln(n+1)-n \ln n\r]\label{gcht}
\ee
The corresponding entropy
\be
S(T,n)&=&(n+1)\ln(n+1)-n \ln n\label{ent}
\ee
can be understood as follows: Consider $N$ (classical) particles to be distributed on $L$ lattice sites. There are
\be
\Omega(N,L)=\frac{(N+L)!}{(N+1)!(L-1)!}\nn
\ee
possibilities of doing this. The associated combinatorial entropy per lattice site is $S_\Omega(N,L)=\frac{1}{L}\ln\Omega(N,L)$. In the thermodynamic limit $N,L\to\infty$, $N/L=n$ fixed, this agrees with \refeq{ent}.

Proceeding further, one finds the linear integral equation for the coefficient $\al_1$:
\be
\al_1(k)  \te^{-\beta \mu_0}&=& \cos k +\int_{-\pi}^\pi \kappa(k-k') \al_1(k')\frac{\d k'}{2\pi}\label{al1},
\ee
and the corresponding expansion of $g$:
\be
g(T,\mu)=T \ln\l(1-\te^{-\beta \mu}\r)-\int_{-\pi}^\pi \al_1(k') \frac{\d k'}{2\pi}\nn .
\ee
From Eq.~\refeq{al1} it follows that $\int_{-\pi}^\pi \al_1(k')\d k'=0$, such that the leading $T$-dependent contribution to $g$ is $\Or\l(\b\r)$. 

In order $\beta^2$, the insertion of Eq.~\refeq{aht} into Eq.~\refeq{aint} yields
\be
\frac{-\al_1^2(k) + 2\al_2(k)\l(\te^{\al_0}-1\r)}{4\cosh^2 \frac{\al_0}{2}}=\int_{-\pi}^\pi \kappa(k-k')\al_2(k')\frac{\d k'}{2\pi}\nn\,.
\ee
Once $\al_2$ is determined from this equation, one has
\be
\frac{g(T,\mu)}{T}= \ln\l(1-\te^{-\beta \mu}\r) -\b^2\int_{-\pi}^\pi \al_2(k') \frac{\d k'}{2\pi}+\Or\l(\beta^{-3}\r)\nn ,
\ee
from which one can calculate $n=-\partial_\mu g(T,\mu)$ and therefrom the $T$-dependence of $\mu$ for fixed $n$, i.e. the coefficient $\mu_1$ in the expansion $\mu(T)=\mu_0+\beta \mu_1$. Eventually, one arrives at the canonical potential $g_c(T,n):=g(T,\mu(T,n))-\mu(T,n) n$, from which one obtains $C(T)|_n$ with the leading $T$-dependence $C(T)\sim \beta^2$. We do not carry out this program here in its full generality, but rather concentrate on the $q=0$-case, where the integral equations can be solved and thus the coefficients of the high-temperature expansion calculated explicitly. 

\subsection{The $q=0$-case} 
For $q=0$, we take advantage of Eq.~\refeq{gdir} in \ref{app1}. By expanding the integrand there in powers of $\beta \cos k$, one obtains
\be
-\b g(T,\mu)&=&\l[\ln\l(1+\te^x\r)+\frac{\b^2}{16 \cosh^2 \frac{x}{2}}+\frac{\b^4}{512}\frac{-2+\cosh x}{\cosh^4\frac{x}{2}}\r.\nn\\
& & \l.+\frac{\b^6}{73728}\frac{33-26\cosh(x)+\cosh 2x}{\cosh^6\frac{x}{2}}\r]_{x=\beta(\mu-g(T,\mu))}\label{gmug}\,.
\ee
Furthermore, also from Eq.~\refeq{gdir}, the particle density is obtained as
\be
n(T,\mu)&=&\frac{\al(T,\mu)}{1-\al(T,\mu)}\nn\\
\al(T,\mu)&=&\int_{-\pi}^\pi\frac{1}{1+\te^{-\beta \cos k -\beta \mu +\beta
    g(T,\mu)}}\frac{\d k}{2\pi}\nn,
\ee
which yields $n$ in terms of $\mu-g(T,\mu)$:
\be
\fl n=\l[\te^x-\frac{\beta^4}{4} \te^x\tanh\frac{x}{2} + \frac{\beta^4}{128}\te^x\frac{3+\cosh x +2 \sinh x}{\cosh^2\frac{x}{2}}\tanh\frac{x}{2} +\frac{\beta^4}{18432} \te^x\r.\nn\\
\fl\; \l.\times\frac{42+16 \cosh x + 10 \cosh(2x) + 54 \sinh x + 9 \sinh(2 x)}{\cosh^4\frac{x}{2}}\tanh\frac{x}{2}\r]_{x=\beta(\mu-g(T,\mu))} \label{nmug}.
\ee
From Eq.~\refeq{nmug}, $\beta(\mu-g)$ is obtained as a function of $n$ order by order in $\beta^2$. This result is then used in Eq.~\refeq{gmug} to express $g$ in terms of $n$. Finally, the first terms of high-temperature-expansions of the canonical potential $g_c$ and the specific heat read
\be
\fl g_c(T,n)= -T\l[(n+1)\ln (n+1)-n \ln n\r] - \beta\frac{n}{4(1+n)} + \beta^4\frac{n(1+n^2)}{64(1+n^3)}\nn\\
\fl \qquad -\beta^6\frac{n(2-n+6 n^2-n^3+2n^4)}{1152(1+n)^5}\nn\\
\fl C(T,n)= \frac{n}{2(1+n)}\beta^2 - \frac{3n(1+n^2)}{16(1+n)^3}\beta^4+\frac{5n(2-n+6n^2-n^3+2n^4)}{192(1+n)^5}\beta^6\label{htc}.
\ee
\subsection{Comparison with numerics for $q=0$}
\begin{figure}
\begin{center}
\includegraphics*[scale=0.5]{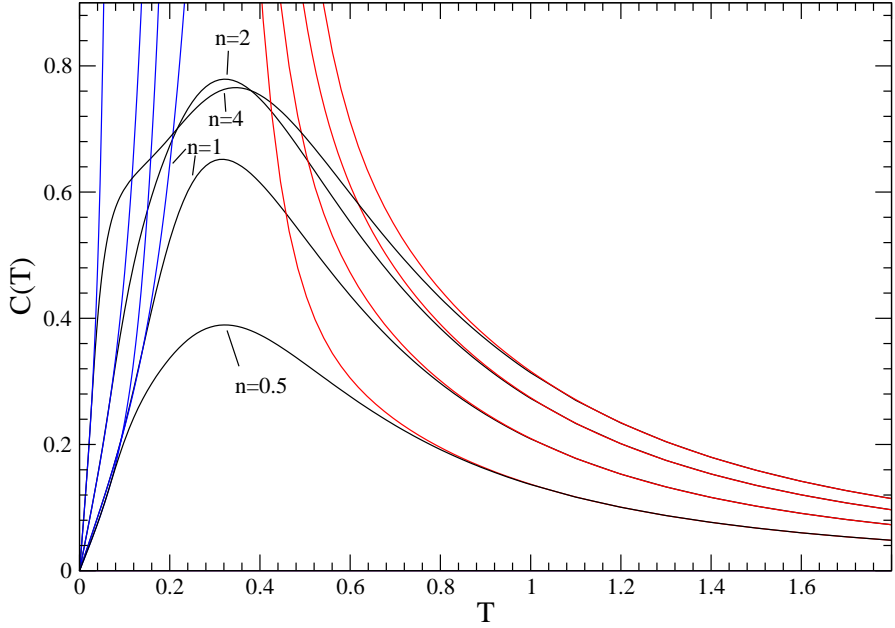}
\caption{The specific heat for $q=0$, $n=0.5,1,2,4$, together with the low- and
  high-temperature asymptotics according to Eqs.~(\ref{sphq0}, \ref{htc}).} 
\label{fig1}
\end{center}
\end{figure}

\begin{figure}
\begin{center}
\includegraphics*[scale=0.5]{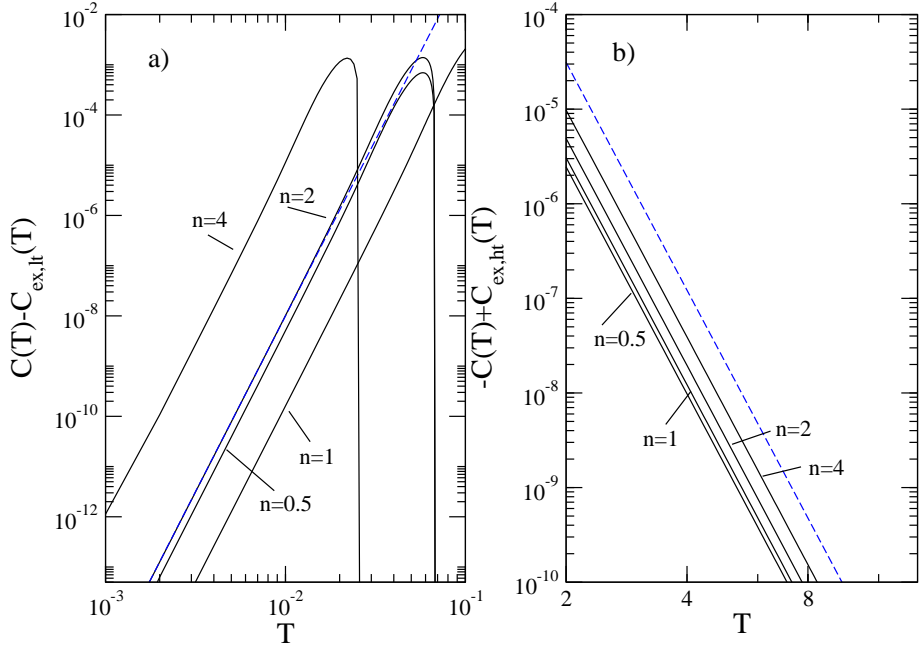}
\vspace{0.2cm}
\caption{Low- and high-temperature asymptotics of the specific heat at $q=0$
  and $n=0.5,1,2,4$. In a), the order $T^7$ of $C(T)$ is shown on a double logarithmic scale,
  after having subtracted the $T$, $T^3$, $T^5$-contributions according to
  Eq.~\refeq{sphq0} (denoted by $C_{ex,lt}$ in the figure). The dashed straight line is $\propto T^7$ and a guideline to the eye. Analogously, in b), the $\beta^8$-contribution in $C$ is
  shown, where the orders $\b^2,\b^4,\b^6$ have been subtracted following Eq.~\refeq{htc} (denoted by $C_{ex,ht}$ in the figure). Here, the straight dashed line is $\propto \beta^8$.} 
\label{fig2}
\end{center}
\end{figure}

The difficulty in a numerical calculation lies in the steep ($\sim \beta$)
gradient of $\ve(k)$ around $k\approx \pm B$. The induced numerical inaccuracy
generally is too high to allow for a quantitative check of the coefficients
(although the exponents in the low-$T$-expansion could be identified,
\cite{borser06}). However, at $q=0$ the iterative numerical approach becomes
particularly accurate because only one non-linear integral equation has to be
solved (see Eq.~\refeq{gdir} in the appendix). Results for different $n$ at
$q=0$ are shown in figures \ref{fig1}, \ref{fig2}. 
\section{Low-temperature asymptotics of the $XXZ$-model in a field}
It turns out that a low-temperature expansion for the free energy of the
spin-1/2 $XXZ$-model in a magnetic field is technically very similar to the
one given above for the $q$-Bose model. This is not too surprising, as argued in the introduction. 

The Hamiltonian is defined as
\be
H=J \sum_{\ell=1}^L \l[S^x_\ell S^x_{\ell+1} + S^y_\ell S^y_{\ell+1} + (\cos \g) S^z_\ell S^z_{\ell+1}\r] +h \sum_{\ell=1}^L S^z_\ell\label{hxxz}\;,
\ee
where $0\leq \g <\pi$ is chosen such that the model is critical and low-energy excitations have a linear dispersion relation. The summation runs over the $L$ lattice sites of the chain, where periodic boundary conditions are imposed. 

Currently, three (equivalent, \cite{takshiak01}) approaches to calculate the thermodynamics exactly are known: the TBA, the QTM and the
single-integral-equation approach \cite{tak01}. For our purposes, the formulation developed within the framework of the QTM-technique is best suited. The free energy per lattice site is given by
\be
f(T,h)&=&e_0-\frac{T}{2\pi} \frac\pi\g\l[d*\ln \B\BB\r](0)\label{fex}\\
\ln \ba(x)&=& -\b v_s^{(0)} d(x) +\frac{\pi}{2(\pi-\g)}\b h +\l[\kappa*\ln \B-\kappa_-\ln \BB\r](x) \nn\\
\ln \bb(x)&=& -\b v_s^{(0)} d(x) -\frac{\pi}{2(\pi-\g)}\b h +\l[\kappa*\ln \B-\kappa_+\ln \BB\r](x) \nn\\
v_s^{(0)}&=&J\frac{\pi\sin\g}{2 \g}\nn\\
\B&=&1+\ba, \; \; \BB=1+\bb\nn\\
d(x)&=&\frac{1}{\cosh\frac\pi\g x}\nn\\
\kappa(x)&=& \i \frac{\sinh\l(\frac\pi2-\g\r)k\,\te^{\rmi k x}}{2\cosh\frac\g2k \, \sinh\l(\frac\pi2-\frac\g2\r)k}\, \frac{\d k}{2\pi}\nn\\
\kappa_\pm (x)&=& \kappa(x\pm \rmi \g\mp \rmi \e)\nn
\ee 
and convolutions are defined as $\l[f*g\r](x):=\i f(x-y)g(y)\d y$. The quantity $e_0$ is the ground-state energy, $v_s^{(0)}$ the spin velocity at zero magnetic field. 
We now focus on the low-temperature regime, that is, $\beta h \gg 1$. In this case, $\ln \BB\sim \exp\l[-\beta h\r]$ is exponentially small and thus is neglected. It is convenient to introduce $\beta \ve(x):=\ln \ba(x)$. Then
\be
\fl f(T,h)-e_0=-\frac{T}{2\pi} \i \frac{\pi}{\g}d(x)\ln\l(1+\te^{\beta \ve(x)}\r)\d x\nn\\
\fl\qquad= -\frac{1}{2\pi} \i\frac{\pi}{\g}d(x)\ve(x)\d x -\frac{T}{2\pi} \i \frac{\pi}{\g}d(x)\ln\l(1+\te^{-\beta \ve(x)}\r)\d x \label{fxxz}\\
\fl\ve(x)=-v_s^{(0)} d(x)+\frac{\pi}{2(\pi-\g)} h +T \i \kappa(x-y)\ln\l(1+\te^{\beta \ve(y)}\r)\d y\nn\\
\fl\qquad=-v_s^{(0)} d(x)+\frac{\pi}{2(\pi-\g)} h \nn\\
\fl\qquad\qquad+\i \kappa(x-y)\ve(y)\d y +T \i \kappa(x-y)\ln\l(1+\te^{-\beta \ve(y)}\r)\d y\nn.
\ee
Using Fourier transform techniques, the last equation is written in a more compact manner as (we rename the spectral parameter as $k$)
\be
\fl\ve(k)=-v_s^{(0)}\frac{\g}{\pi} \hat\kappa_{\frac\g2}(k)+\beta h +\i\hat\kappa_\g(k-k') \ln \l(1+\te^{-\beta \ve(k')}\r)\frac{\d k'}{2\pi}\label{exxz}\\
\fl\hat\kappa_\g(k)=\frac{2\sin(2\g)}{\cosh(2 k)-\cos(2\g)}\nn,
\ee
which makes the similarity with Eq.~\refeq{vefint} apparent. 

One now goes through the same steps as in the analysis above. We give the corresponding results in the following. The functions $\rho_{0,1}$ satisfy the linear integral equations
\be
\rho_\nu(k)&=&\rho^{(\nu)}(k) + \int_{-B_0}^{B_0} \hat\kappa_\g(k-k') \rho_\nu(k')\frac{\d k'}{2\pi}\nn\\
\rho^{(0)}(k)&=&\frac{1}{2\g}d(k)\nn\\
\rho^{(1)}(k)&=&B_1\frac{\rho_0}{2\pi}\l[\hat\kappa_\g(k-B_0)+\hat\kappa_\g(k+B_0)\r]\nn
\ee
and for the functions $\ve_{0,1}(k)$, we find
\numparts
\label{vexxz}
\be
\fl\ve_\nu(k)=\ve^{(\nu)}(k) + \int_{-B_0}^{B_0} \hat\kappa_\g(k-k') \ve_\nu(k')\frac{\d k'}{2\pi}\\
\fl\ve^{(0)}(k)=-v_s^{(0)}\frac{\g}{\pi} \hat\kappa_{\frac\g2}(k)+ h\\
\fl\ve^{(1)}(k)=-\frac{\pi}{12\ve_0'}\l[\hat\kappa_\g(k-B_0)+\hat\kappa_\g(k+B_0)\r] +\int_{-B_0}^{B_0}\hat\kappa_\g(k-k')\ve_1(k')\frac{\d k'}{2\pi}.
\ee
\endnumparts
The shift $B_1$ is given by Eq.~\refeq{b1}, with $\ve_{1,0}$ specified in Eqs.~\refeq{vexxz}. The low-temperature expansion of the free energy reads
\numparts
\label{fxxzlt}
\be
f(T,h)&=&f_0(h)+f_1(h)T^2+f_2(h)T^4+\Or\l(T^6\r)\label{fxxz1}\\
f_0(h)&=&e_0+\frac{1}{2\pi} \int_{|k|>B_0}d(k)v_s^{(0)} \hat\kappa_{\frac\g2}(k)   \d k\\
f_1(h)&=&-\frac{\pi^2}{3}\frac{\rho_0}{\ve'_0}\\
f_2(h)&=&-\l\{\frac{7\pi^4}{15}\l[\rho_0\l(\frac{(\ve_0'')^2}{4\ve_0'^5}-\frac{\ve_0^{(3)}}{12\ve_0'^4}\r)-\frac{\rho_0'\ve_0''}{4\ve_0'^4}+\frac{\rho_0''}{12\ve_0'^3}\r]\r.\nn\\
& &\l.+\frac{\rho_0\ve_1^2}{\ve_0'}+\frac{\pi^2}{3}\l[-\frac{\ve_1'\rho_0}{\ve_0'^2}-\frac{\rho_0'\ve_1}{\ve_0'^2}+\frac{\ve_0''\rho_0\ve_1}{\ve_0'^3}\r]\r\}\label{fxxzl}
\ee
\endnumparts
where $\ve_0:=\ve_0(B_0)$ etc. 

In this model, the particle density is constant per construction, so that the specific heat can be calculated directly from $C(T,h)/T=-\partial_T f(T,h)$. However, formula \refeq{ct}, with $\mu$ replaced by $h$, can be used to calculate the specific heat at constant magnetization $m$. The results in Eqs.~(\ref{fxxz1}-\ref{fxxzl}) are confirmed for free fermions, cf. \ref{app2}.

\section{Summary and outlook}
The $q$-Bose model and the critical spin-1/2 Heisenberg chain in an external
magnetic field were studied in the regime of low temperatures. In both cases,
the thermodynamic potentials could be expanded in terms of even powers of
$T$. We developed a method that can be used to calculate the coefficients of
this expansion, and carried out the calculation for the $T^4$-contribution. 

Directions for future research are twofold: One question is in how far the
results obtained here can be used to determine the coefficients of the leading
irrelevant operators in the low-energy effective Hamiltonians for these
models. The other question concerns the relationship between the spin chain
and the bosonic model. The absence of backscattering in both models (at low energies for the spin chain) suggests
that a relationship exists on a more fundamental level. 
\section*{Acknowledgments}
The author thanks M. T. Batchelor, X.-W. Guan, S. Sergeev and
J. Sirker for helpful discussions. This work has been supported by the German Research Council (DFG) under grant number BO 2538/1-1.
\appendix
\section{Specific heat at $q=0$}
\label{app1}
By combining Eqs.~(\ref{g}, \ref{vefint}), it is not difficult to show that for $q=0$,
\be
g(T,\mu)=-T\int_{-\pi}^\pi \ln\l(1+\te^{\beta\cos k+\beta\mu-\beta g(T,\mu)}\r)\frac{\d k}{2\pi}\label{gdir}.
\ee
In the low-temperature regime $\beta\gg |\mu|$, 
\be
g(T,\mu)=-\beta \int_{-\pi}^\pi \frac{(-k \cos k + \sin k)\sin k}{4\cosh^2\frac\beta2(\cos k + \mu-g(T,\mu))}\frac{\d k}{2\pi}\nn,
\ee
which is obtained by two integrations by parts, neglecting exponentially small quantities. One now substitutes $u=\beta (\cos k+\mu-g(T,\mu))$ and obtains
\be
\fl g(T,\mu)=-2 \i \frac{\sqrt{1-\l(\frac{u}{\beta}-(\mu-g)\r)^2}-\l(\frac{u}{\beta}-(\mu-g)\r)\arccos\l[\frac{u}{\beta}-(\mu-g)\r]}{4\cosh^2\frac{u}{2}}\frac{\d u}{2\pi}\nn,
\ee
where the integration boundaries have been sent to $\pm \infty$, again
neglecting exponentially small corrections, and the $(T,\mu)$-dependence of
$g$ has not been displayed explicitly on the rhs. One now expands the integrand in powers of $u/\beta$ and makes use of Eqs.~(\ref{zeta1}, \ref{zeta2}), resulting in 
\be
g(T,\mu)&=& -\frac{1}{\pi}\l[\sqrt{1-(\mu-g)^2}+(\mu-g)\arccos(g-\mu)\r]\label{gq0}\\
& &-T^2\frac{\pi}{6}\frac{1}{\sqrt{1-(\mu-g)^2}}\nn\\
& &-T^4 \frac{7\pi^3}{360} \frac{1+2(\mu-g)^2}{(1-(\mu-g)^2)^{5/2}}\label{gbare}\\
& &-T^6 \frac{31 \pi^5}{5040}\frac{3+8(\mu-g)^2(3+(\mu-g)^2)}{(1-(\mu-g)^2)^{9/2}}+\Or\l(T^8\r)\nn.
\ee
This equation for $g$ can be solved order by order by expanding $g=g_0+g_1T^2+g_2T^4+g_3T^6$. This allows for the calculation of $n(\mu,T)$, and therefrom, by holding $n$ fixed, the $(T,n)$-dependence of $\mu$ is derived: 
\be
\fl \mu(T,n)=-\l(\frac{\cos B_0}{n+1}+\frac1\pi\sin B_0\r)\nn\\
\fl\qquad +\frac{\pi\l(-1-n+\pi\cot B_0\r)}{6(n+1)\sin B_0}T^2\nn\\
\fl\qquad  +\frac{\pi^3\l(293\pi\cos B_0+27\pi \cos (3 B_0)-(n+1)\l(37 \sin B_0+9\sin (3B_0)\r)\r)}{1440 (n+1) \sin^6 B_0}T^4\label{mutq0}\\
\fl\qquad  + \frac{\pi^5}{725760 \sin^{10} B_0}\l(1088882\pi \cos B_0 + 275885\pi \cos (3 B_0) + 11153\pi \cos (5B_0)\r.\nn\\
\fl\qquad\qquad \l.-(n+1)(38354 \sin B_0 + 26487 \sin (3 B_0) + 1525 \sin (5 B_0))\r)T^6\nn\\
\fl B_0= \frac{n\pi}{n+1}\label{b0}.
\ee
Taking this $T$-dependence of $\mu$ into account, one arrives at the specific heat, taken at constant $n$, Eq.~\refeq{sphq0}. Furthermore, the charge susceptibility is obtained as $\chi_c(T,n)=1/\partial_n \mu(T,n)$. 

Let us now confirm these results from the general formulae (\ref{ggen}-\ref{g2}, \ref{ctn}). Eqs.~\refeq{rhonuex} (for $\nu=0$ there) and \refeq{vet0} are solved by
\be
\rho_0&=&\frac{1}{2(\pi-B_0)}\nn\\
\ve_0(k)&=&-\cos k -\mu -\frac{\sin B_0 + B_0 \mu}{\pi-B_0}\nn.
\ee
The condition $\ve_0(\pm B_0)=0$ yields 
\be
\mu=-\frac{\sin B_0}{\pi} +\l(\frac{B_0}{\pi}-1\r)\cos B_0\nn
\ee
According to Eqs.~(\ref{vefint}, \ref{gq0}), $\ve(k)=-\cos k -\mu+g$, and thus $\cos B_0=g_0-\mu$. On the
other hand, we know \cite{borser06} that $\int_{-B_0}^{B_0} \rho_0 \d
k/2\pi=n$, which yields Eq.~\refeq{b0}. This confirms the $T^0$-order of
Eq.~\refeq{mutq0} and is consistent with $n(\mu_0)=-\partial_{\mu_0}
g_0(\mu_0)$. However for the sake of brevity, we do not display the $\mu_0$-dependence
of $n$. The coefficients (\ref{g1ex}, \ref{g2}) can thus be calculated and read
\be
g_1&=&-\frac{\pi}{6 v_c}\nn\\
g_2&=& \frac{\pi}{\pi-B_0}\l(-\frac{\pi}{72v_c^2\sin B_0}+\frac{\pi^2\cos B_0}{36 v_c\sin^3B_0}-\frac{7\pi^3(1+2\cos^2 B_0)}{360\sin^5 B_0}\r)\nn\\
v_c&=&\frac{1}{n+1} \,\sin\frac{\pi n}{n+1}\nn.
\ee
These results are confirmed by the slightly different analytical calculation sketched in the first part of this appendix. Note that the $q=0$-case is special in so far as $\rho_0(k)$ is a constant, as well as $\ve_1(k)$. 

\section{Specific heat for free fermions}
\label{app2}
We calculate the first terms of a low-temperature expansion of the free energy for free spinless fermions on a lattice in two different ways: Directly and using the exact solution. Both are equivalent and yield the same results. 

At $\g=\pi/2$, a Jordan-Wigner transformation followed by a Fourier transformation of the Hamiltonian \refeq{hxxz} yields
\be
H=\sum_{j=1}^N(J \cos k_j+h)c^\dagger_{k_j}c_{k_j}\qquad \mbox{with  } k_j=\frac{2\pi}{N}j\nn.
\ee
Thus in the thermodynamic limit, 
\be
-\beta f(T,h)=\frac{1}{2\pi} \int_{0}^{2\pi} \ln\l(1+\te^{-\beta(J\cos x+h)}\r)\d x\nn,
\ee
which is very similar to Eq.~\refeq{gdir}. Proceeding analogously to there, one finds
\be
f(T,h)&=&-\frac{J}{\pi}\l(\sqrt{1-\l(\frac{h}{J}\r)^2}-\frac{h}{J}\arccos\frac{h}{J}\r)\nn\\
& & -J \frac\pi6 \l(\frac{T}{J}\r)^2\frac{1}{\sqrt{1-(h/J)^2}}\label{fflim}\\
& & -J \frac{7 \pi^3}{360} \l(\frac{T}{J}\r)^4\frac{1+2(h/J)^2}{\sqrt{1-(h/J)^2}^{5/2}}\nn.
\ee
On the other hand, we may use the exact solution Eq.~\refeq{fxxzlt} with 
\be
\rho(k)&=&\frac{1}{\pi\cosh 2 x}\nn\\ 
\ve(k)&=& -\frac{J}{\cosh 2x}+h\nn\\
B&=&\frac12\mbox{arccosh}\frac{J}{h}\nn.
\ee
Note that $\rho(k)\equiv\rho_0(k),\;\ve(k)\equiv\ve_0(k),\;B\equiv B_0$ here. Inserting these values into Eq.~\refeq{fxxzlt}, one confirms the result \refeq{fflim}.

\end{document}